\documentclass[aps,prb,twocolumn,groupedaddress,showpacs,floatfix]{revtex4}

\usepackage{amsmath,amssymb,graphicx,bm,dcolumn}

\begin{document}

% \preprint{}

\title{
On the Efficiency of Topological Defect Formation \\
in the Systems of Various Size and (Quasi-) Dimensionality
}

\author{Yurii V.\ Dumin}
\altaffiliation[Also at ]{%
Institute for Pure and Applied Mathematics (IPAM),
University of California, Los Angeles (UCLA),
Box 957121, Los Angeles, CA 90095-7121, USA}
\email{dumin@yahoo.com}
\affiliation{%
Max-Planck-Institut fuer Physik komplexer Systeme,\\
Noethnitzer Strasse 38, 01187 Dresden, Germany}

\author{Ludmila M.\ Svirskaya}
\email{svirsk@cspi.urc.ac.ru}
\affiliation{%
Chelyabinsk State Pedagogical University,\\
prosp.~Lenina~69, 454048 Chelyabinsk, Russia}

\date{March 29, 2005}

\begin{abstract}
The experiments on verification of the Kibble--Zurek mechanism
showed that topological defects are formed most efficiently
in the systems of small size or low (quasi-)dimensionality,
whereas in the macroscopic two- and three-dimensional samples
a concentration of the defects, as a rule, is strongly suppressed.
A reason for universality of such behavior can be revealed by
considering a strongly-nonequilibrium symmetry-breaking phase transition
in the simplest ${\varphi}^{4}$ field model. It is shown that
the resulting distribution of the defects (domain walls)
is formally reduced to the well-known Ising model,
whose behavior changes dramatically in passing from
a finite to infinite size of the system and
from the low (D=1) to higher (D$\ge$2) dimensionality.
\end{abstract}

% 05.70.Fh - Phase transitions: general studies
% 64.60.Ht - Dynamic critical phenomena
% 64.60.Cn - Order-disorder transformations; statistical mechanics of model systems
% 03.75.Lm - Tunneling, Josephson effect, Bose-Einstein condensates in periodic
%            potentials, solitons, vortices and topological excitations
% 03.75.Gg - Entanglement and decoherence in Bose-Einstein condensates

%\keywords{Suggested keywords}% Use showkeys class option if keyword
                              % display desired
\maketitle

\section{Introduction}

\subsection{\label{sec:ConTopDef}
The concept of topological defects}

Formation of topological defects by the strongly-nonequilibrium
symmetry-breaking phase transitions is the subject of interest
both in condensed matter and field theory.
This is because of a close similarity between the Lagrangian
of Landau--Ginzburg theory, widely used to describe phase transitions
in the condensed-matter systems, and the Lagrangians of
the modern elementary-particle theories (such as the standard
electroweak model or various kinds of the Grand Unification Theories),
which are substantially based on the concept of spontaneous
symmetry breaking.

After the phase transitions in all the above-mentioned cases,
the stable topological defects of the order parameter can arise,
such as the monopoles, strings (vortices), and domain walls,
depending on the symmetry group involved.

The most efficient mechanism of the defect formation is
the so-called Kibble--Zurek scenario,\cite{Kibble76,Zurek85} which is based
on the simple causality arguments. Namely, if during the phase transition
the information about the order parameter can spread over
the distance ${\xi}_\text{eff}$, then the phases of the order parameter
should be established independently in the regions of characteristic
size ${\xi}_\text{eff}$.%
\footnote{The effective correlation length ${\xi}_\text{eff}$,
appearing in the Kibble--Zurek scenario, should not be mixed with
the coherence length commonly introduced in the Landau--Ginzburg theory
for the systems in thermodynamic equilibrium or quasi-equilibrium.}
As a result, after some relaxation following the phase transition,
the stable defects of the order parameter should be formed
at the typical separation ${\xi}_\text{eff}$ from each other.
So, their concentration in a 3-dimensional system
can be roughly estimated as
\begin{equation}
n \approx 1 / \, {\xi}_\text{eff}^{\,d},
\end{equation}
where $ d = $~3, 2, and 1 for the monopoles, strings (vortices),
and domain walls, respectively;
while the effective correlation length ${\xi}_\text{eff}$
depends on the particular system under consideration.

For example, in the symmetry breaking of Higgs fields
by the cosmological phase transitions (i.e.\ a generation of mass
of the elementary particles) ${\xi}_\text{eff}$ is
commonly taken to be\cite{Klapdor97}
\begin{equation}
{\xi}_\text{eff} \, \lesssim \, c \, / H_\text{pt} \, ,
\end{equation}
where $ c $ is the speed of light, and
$ H_\text{pt} $ is Hubble constant at the instant of phase transition.

In the consideration of vortex generation by
a superfluid phase transition in helium,
the corresponding quantity is
\begin{equation}
{\xi}_\text{eff} \, \approx \, c_2 \, {\tau}_{\scriptscriptstyle Q} \, ,
\end{equation}
where $ c_2 $ is the speed of the second sound
(i.e. a characteristic speed of propagation of information about
the phase of the order parameter),
and $ {\tau}_{\scriptscriptstyle Q} $ is the so-called quench time
(i.e. a characteristic time of the phase transition),
defined as
\begin{equation}
{\tau}_{\scriptscriptstyle Q}^{-1} =
  \, \left. \left( \frac{1}{T} \frac{dT}{dt} \right) \right| _{T = T_c} .
\end{equation}

The similar definitions of ${\xi}_\text{eff}$ are used also for other
systems (superconductors, liquid crystals, etc.).

\subsection{Review of experimental data}

Although Kibble--Zurek scenario was proposed initially
in the context of cosmological phase transitions
in the field theories admitting the symmetry breaking
(in fact, the first idea of such kind was put forward by
Bogoliubov\cite{Bogoliubov66} almost 40 years ago),
much efforts have been undertaken in the last decade
to verify this phenomenon by the laboratory experiments.
These works were originated by
Chuang, \textit{et~al.}\cite{Chuang91} in 1991,
and about a dozen of such experiments were performed by now
in various superfluid and superconducting systems.
They are summarized in Table~\ref{tab:SumExp}.
(This table does not show a number of experiments with liquid crystals,
aimed at the detailed study of inner structure of the defects
without measuring the rates of their formation.)

%%%%%%%%%%%%%%%%%%%%%%%%%%%%%
%%%%%%%%%% TABLE 1 %%%%%%%%%%
%%%%%%%%%%%%%%%%%%%%%%%%%%%%%
%
\begin{table*}
\caption{\label{tab:SumExp}
Summary of experiments on the defect formation by
strongly-nonequilibrium symmetry-breaking phase transitions.
A positive result~($+$) implies an agreement
with the Kibble--Zurek estimate within a factor about unity;
and the negative result~($-$), a disagreement by
a few times to a few orders of magnitude.
}
\begin{ruledtabular}
\begin{tabular}{lllllcccc}
\multicolumn{3}{l}{} & \multicolumn{1}{c}{\textbf{Initiation}} &
\multicolumn{1}{c}{\textbf{Method}} &   \textbf{(Quasi-)} &
\textbf{Size} & \textbf{Result}
\\
\multicolumn{3}{l}{\textbf{Experimental object}} &
\multicolumn{1}{c}{\textbf{of phase}} &
\multicolumn{1}{c}{\textbf{of}} & \textbf{dimen-} & \textbf{of} &
\textbf{of expe-}& \textbf{Refs.}\footnotemark[1]
\\
\multicolumn{3}{l}{} & \multicolumn{1}{c}{\textbf{transition}} &
\multicolumn{1}{c}{\textbf{detection}} & \textbf{sionality} &
\textbf{sample} & \textbf{riment}
\\
\hline
\\
%%%%%%%%%% SUPERFLUIDS %%%%%%%%%%
%
\multicolumn{2}{l}{Superfluids} & ${}^4$He & Expansion &
Second sound & 3 & macro & \textbf{--} &
\onlinecite{Hendry94,Dodd98}\footnotemark[2]
\\
\multicolumn{2}{l}{} & & of sample & absorption
\\
\multicolumn{2}{l}{} & ${}^3$He & Neutron & Calorimetry & 3 & micro &
\textbf{+} & \onlinecite{Bauerle96}
\\
\multicolumn{2}{l}{} & & irradiation
\\
\multicolumn{2}{l}{} & & & Nuclear magne- & 3 & micro & \textbf{+} &
\onlinecite{Ruutu96}
\\
\multicolumn{2}{l}{} & & & tic resonance
\\
%%%%%%%%%% SUPERCONDUCTORS %%%%%%%%%%
%
Super- & \multicolumn{2}{l}{Thin films} & Heating-- & SQUID & 2 & macro &
\textbf{--} & \onlinecite{Carmi99,Maniv03}
\\
conduc- & & & cooling
\\
tors & & & cycles & SQUID  micro-& 2 & macro & \textbf{--} &
\onlinecite{Kirtley03}
\\
& & & & scope
\\
& \multicolumn{2}{l}{Multi-Josephson-} & & SQUID & 1 & macro & \textbf{+} &
\onlinecite{Carmi00}
\\
& \multicolumn{2}{l}{junction loop}
\\
& \multicolumn{2}{l}{Annular Josephson} & & Voltage mea- & 1 & macro &
\textbf{+} & \onlinecite{Monaco02}
\\
& \multicolumn{2}{l}{tunnel junctions} & & surement
\\
\end{tabular}
\end{ruledtabular}
\footnotetext[1]{Only the first publication by each experimental group
is shown.}
\footnotetext[2]{The initial results of Ref.~\onlinecite{Hendry94}
were subsequently corrected in Ref.~\onlinecite{Dodd98}.}
\end{table*}
%
%%%%%%%%%%%%%%%%%%%%%%%%%%%%%
%%%%%%%%%%%%%%%%%%%%%%%%%%%%%
%%%%%%%%%%%%%%%%%%%%%%%%%%%%%

Analysis of the table reveals a quite interesting tendency:
\textit{the topological defects are formed most efficiently in the systems
of small size or low (quasi-) dimensionality,}\footnote{
The term quasi-dimensionality implies here that thickness of the sample
in some direction(s) is so small that it can be ignored, and this sample
can be considered as the system of less effective dimensionality.}
for example,
the one-dimensional multi-Josephson-junction loops (MJJL)~\cite{Carmi00}
and annular Josephson tunnel junctions~\cite{Monaco02}
as well as in microscopic hot bubbles of ${}^3$He
produced by neutron irradiation.~\cite{Bauerle96, Ruutu96}
On the other hand, \textit{the concentration of the defects in
macroscopic systems of higher dimensionality,} e.g.\ the two-dimensional
superconductor films~\cite{Carmi99, Maniv03, Kirtley03} and three-dimensional
volume samples of ${}^4$He (Refs.~\onlinecite{Hendry94, Dodd98}),
\textit{was found to be considerably less than the theoretical predictions.}

In principle, there may be various explanations of the above fact
(for example, different material parameters of the substances
listed in Table~\ref{tab:SumExp}).
The aim of the present work is to propose another explanation,
which is based on the universal geometric properties
of the systems under consideration, namely,
their size and dimensionality (or quasi-dimensionality).

\section{The model of defect formation}

\subsection{Initial assumptions and equations}

To carry out all calculations in analytic form, let us consider
the simplest ${\varphi}^{4}$-model of real scalar field
(the order parameter), whose Lagrangian
\begin{equation}
{\cal L} \, ({\bf r}, t) \, = \,
\frac{1}{2} \, \big[ {\left( {\partial}_t \varphi \right)}^2 \! - \,
  {\left( \nabla \varphi \right)}^2 \big]
  \, - \:
\frac{\lambda}{4} \,
  {\big[ \, {\varphi}^2 \! - \left( {\mu}^2 / \lambda \right) \big]}^2
\label{Lagrangian}
\end{equation}
admits the discrete ($ \mathbb{Z}_2 $) symmetry breaking.

Two stable vacuum states of this field (which, for the sake of convenience,
will be marked by the oppositely directed arrows) are
\begin{equation}
{\varphi}_{\uparrow \downarrow} =
\pm \, {\varphi}_0 =
\, \pm \, \mu \, / \sqrt{\lambda} \; ;
\label{VacState}
\end{equation}
the structure of a domain wall between them, located at $ x = x_0 $,
is described as
\begin{equation}
\varphi \, (x) = \, \pm \, {\varphi}_0
  \tanh \big[ \frac{\mu}{\sqrt{2}} \,
  (x - x_0) \big] \, ;
\end{equation}
and the specific energy concentrated in this wall equals
\begin{equation}
E = \, \frac{2 \, \sqrt{2}}{3} \, \frac{{\mu}^3}{\lambda} \; .
\end{equation}

Let a domain structure formed after a strongly-nonequilibrium
phase transition in this model be approximated by
a regular rectangular grid with a cell size about the effective
correlation length $ {\xi}_\text{eff} $,
whose definition in the Kibble--Zurek scenario was already discussed
above in Sec.~\ref{sec:ConTopDef}.
The particular value of $ {\xi}_\text{eff} $ is not
of importance here, but we shall assume that it is sufficiently large
in comparison with a characteristic thickness of
the domain wall~$ \sim \! 1 / \mu$.
As a result, the final pattern of vacuum states~(\ref{VacState})
after the phase transition will look like a distribution of spins
on the rectangular grid.

The key assumption of the Kibble--Zurek mechanism is that
the final symmetry-broken states of the field $\varphi$
in two neighboring cells are completely independent of each other.
Then, the probability of formation of a domain wall between them
is given evidently by the ratio of the number of statistical
configurations involving the domain wall to the total number of
configurations:
\begin{equation}
P_{\rm \scriptscriptstyle KZ} = 2/4 = 1/2 \; ;
\end{equation}
and the resulting concentration of the defects (domain walls) will be
\begin{equation}
n_{\rm \scriptscriptstyle KZ} =
  \frac{1}{2} \:
  \frac{\text{D}}{{\xi}_{\text{eff}}^{\text{D}}} \; ,
\end{equation}
where D is the effective dimensionality of the system.

In fact, such approach is not sufficiently accurate, because
a nonzero value of the order parameter formed after the phase transition
represents a coherent state of Bose condensate,
in which the specific quantum correlations may occur even at the distances
exceeding the effective correlation length $ {\xi}_\text{eff} $
derived from the ``classical'' arguments.
This kind of correlations was clearly demonstrated by
the MJJL experiment~\cite{Carmi00}.

\subsection{Overview of MJJL experiment}

\begin{figure*}
\includegraphics[width=16cm]{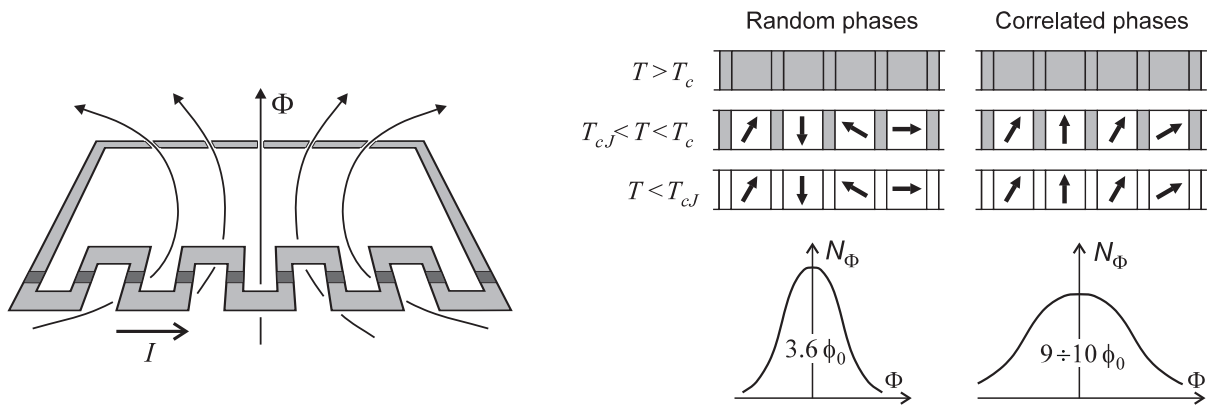}
\caption{\label{fig:MJJL}
Sketch of the experimental setup (left) and basic results (right)
of MJJL experiment.}
\end{figure*}

For the sake of completeness, let us briefly remind
the basic idea and result of the multi-Josephson-junction loop experiment
by Carmi, \textit{et al.,}~\cite{Carmi00} whose sketch is presented
in Fig.~\ref{fig:MJJL}.

%%%%%%%%%% FIGURE 1 %%%%%%%%%%
%%%%%%%%%%%%%%%%%%%%%%%%%%%%%%

A thin quasi-one-dimensional winding strip was engraved at the boundary
between two crystalline grains of YBa${}_2$Cu${}_3$O${}_{7\mbox{-}\delta}$
high-temperature superconductor film, thereby forming
a loop of 214 superconductor segments separated by the grain-boundary
Josephson junctions. This system experienced multiple heating--cooling
cycles in the temperature range 77~K to $\sim$100~K,
which covers both the critical temperature of superconducting phase
transition in the segments of the loop ($T_c = 90$~K)
and in the junctions between them ($T_{cJ} = 83 \div 85$~K).

There is evidently no order parameter in the entire loop as long as
$ T > T_c $. Next, when the temperature drops below~$T_c$ but remains
above~$T_{cJ}$ (i.e. $ T_{cJ} < T < T_c $), some value of the order
parameter should be established in each segment, as is schematically
shown by arrows in the right-hand part of Fig.~\ref{fig:MJJL}.
Since these segments are separated by nonconducting Josephson junctions,
it seems to be reasonable to assume that the phase jumps between them
are random (i.e. uncorrelated to each other).
Finally, when the temperature drops below $T_{cJ}$, the entire loop
becomes superconducting and, due to the above-mentioned jumps,
a phase integral along the loop, in general, should be nonzero.
As a result, the electric current $I$, circulating along the loop,
and the corresponding magnetic flux $\Phi$, penetrating the loop,
will be spontaneously generated.

So, if the phase jumps in the intermediate state $ T_{cJ} < T < T_c $
were absolutely uncorrelated, then distribution of
the spontaneously-trapped magnetic flux in the particular experimental
setup~\cite{Carmi00} would be given by the normal (Gaussian) law
with a characteristic width $3.6\,{\phi}_0$ (where ${\phi}_0$ is
the magnetic-flux quantum). On the other hand, the actual experimental
distribution was found to be over two times wider;
and this anomaly was satisfactorily interpreted by the authors of
the experiment assuming that the phase jumps in the intermediate state
were not random but correlated to each other so that
the probability $P({\delta}_i)$ of the phase difference ${\delta}_i$
in the $i$'th junction was
\begin{equation}
P({\delta}_i) \sim \exp [-{E_J}({\delta}_i) \,/\, T_{c}] \: ,
\label{MJJLcorrel}
\end{equation}
where $E_J$ is the energy concentrated in the Josephson junction, and
$T_{c}$ is the phase transition temperature, measured in energy units.

The nature of the above-mentioned correlations between the phases of
order parameter in the spatially-separated regions is not sufficiently
understood yet. They may be one of manifestations of the specific quantum
``entanglement'', revealed by now in various quantum systems.
We are not going to discuss here this problem in more detail
but assume that the correlations like~(\ref{MJJLcorrel}) are typical for
all Bose condensates formed by the symmetry-breaking phase transitions.
As will be seen from the subsequent analysis, such an assumption
will enable us to interpret satisfactorily a wide range of
experimental data on the efficiency of formation of topological defects.
Just this fact can be considered as indirect confirmation of universality
of~(\ref{MJJLcorrel}).

\subsection{Improvement of the standard estimates}

As follows from the above discussion, the probabilities of various field
configurations after a symmetry-breaking phase transition should be
calculated taking into account the energy concentrated in the defects.
As a result, a pattern of the domains formed by the strongly-nonequilibrium
phase transition in the $ {\varphi}^4 $ lattice model will look like
a distribution of spins in Ising model at the temperature $T$ which is
formally equivalent to the critical temperature $ T_c $ of the initial
$ {\varphi}^4 $-model. All aspects of this formal correspondence are
summarized in Table~\ref{tab:FormCor}.

%%%%%%%%%% TABLE 2 %%%%%%%%%%
%
\begin{table}
\caption{\label{tab:FormCor}
Formal correspondence between the phase transitions in
${\varphi}^4$ and Ising models.
}
\begin{ruledtabular}
\begin{tabular}{rll}
& \multicolumn{1}{c}{$\bm{{\varphi}^4}$\textbf{-model}} &
  \multicolumn{1}{c}{\textbf{Ising model}}
\\
\hline
1. & State formed after a        & Thermodynamically-             \\
   & strongly-nonequilibrium     & equilibrium state.             \\
   & phase transition.           &                                \\
2. & Elementary domains of       & Spins.                         \\
   & the symmetry-broken state.  &                                \\
3. & Statistical sum $Z$ for     & Thermodynamical statistical    \\
   & distribution of the domains & sum $Z$ for spin distribution. \\
   & with various vacuum states. &                                \\
4. & Critical temperatures $T_c$ & Temperature $T$ varying in     \\
   & for various systems.        & the course of the phase        \\
   &                             & transition.                    \\
5. & Suppression of the domain   & Phase transition in the spin   \\
   & wall formation.             & system.                        \\
\end{tabular}
\end{ruledtabular}
\end{table}
%
%%%%%%%%%%%%%%%%%%%%%%%%%%%%%

Then, the probability of formation of a domain wall should be calculated as
\begin{equation}
P = \frac{\hphantom{1}T_{c}^2}{E \, \text{D} N^{\text{D}}} \;
    \frac{\partial}{\partial T_{c}} \ln Z^{(\text{D})} \: ,
\label{gen_def-con}
\end{equation}
where $E$ is the energy of the elementary domain wall
(i.e. at one boundary between two neighboring cells),
$ N $ is the number of cells along each side of the lattice, and
\begin{equation}
Z^{(\text{D})} = \sum_i \exp (- {\varepsilon}_i \, / \, T_{c})
\label{gen_stat-sum}
\end{equation}
is the usual statistical sum over all possible spin configurations of
the Ising model, where $ {\varepsilon}_i $ is the total energy of
$i$'th configuration.

Particularly, for 1-dimensional system with periodic boundary conditions,
\begin{equation}
Z^{(1)} \, = \,
  \sum_{i=1}^{\scriptstyle N}
  \sum_{s_i=\pm 1}
  \exp \left\lbrace \vphantom{\frac{A}{A}} \right.
  \! - \frac{E}{T_{c}} \,
  \sum_{k=1}^{\scriptstyle N} \,
  \frac{1}{2} \, (1 - s_{k} s_{k+1})
  \left. \vphantom{\frac{A}{A}} \right\rbrace ;
\label{stat-sum_1D}
\end{equation}
\smallskip
for 2-dimensional system,
\begin{eqnarray}
Z^{(2)} \, = \,
  \sum_{i=1}^{\scriptstyle N}
  \sum_{j=1}^{\scriptstyle N}
  \sum_{s_{ij}=\pm 1}
  \exp \left\lbrace \vphantom{\frac{A}{A}} \right.
  \! - \frac{E}{T_{c}} \,
  \sum_{k=1}^{\scriptstyle N} \,
  \sum_{l=1}^{\scriptstyle N} \,
  \frac{1}{2} \, ( \, 2 -
\nonumber \\
  s_{kl} s_{k+1, l} \, - \, s_{kl} s_{k, l+1})
  \left. \vphantom{\frac{A}{A}} \right\rbrace ;
\label{stat-sum_2D}
\end{eqnarray}
and so on. Here, $ s_k $ and $ s_{kl} $ are the spin-like variables
describing the broken-symmetry states in the $k$'th and $(kl)$'th cell,
respectively.

As is known (e.g.\ Ref.~\onlinecite{Rumer80}),
\textit{the Ising model for one-dimensional as well as the finite-size
higher-dimensional systems does not experience a phase transition
to the ordered state at any value of the ratio $E/T$.}
From the viewpoint of the domain wall formation by
the strongly-nonequilibrium phase transition in ${\varphi}^{4}$-model
(Table~\ref{tab:FormCor}),
\textit{this means that concentration of the defects will not differ
considerably from the standard Kibble--Zurek estimate,}
because the probability of defect formation $P$ at the scale of
the effective correlation length $ {\xi}_\text{eff} $ will not
deviate substantially from $ P_{\rm \scriptscriptstyle KZ} = 1/2 \sim 1$.

On the other hand, \textit{the Ising model for the sufficiently large
(infinite-size) two- and three-dimensional systems does experience
a phase transition to the ordered state at some value of $E/T \! \sim \! 1$.
As a result, the concentration of domain walls in the corresponding
${\varphi}^{4}$-model at large ratios $E/T_{c}$ should be suppressed
dramatically,} due to formation of macroscopic regions with the same
value of the order parameter, covering a great number of cells of
the effective correlation length~$ {\xi}_\text{eff} $.
(To avoid misunderstanding, let us emphasize once again that
the different values of $T_{c}$ should be understood here as
the critical temperatures for various physical systems described by
the ${\varphi}^{4}$-model, and they formally correspond to
the variable temperature $T$ of a fixed Ising system.)

\subsection{Particular example}

\begin{figure}
\includegraphics[width=7cm]{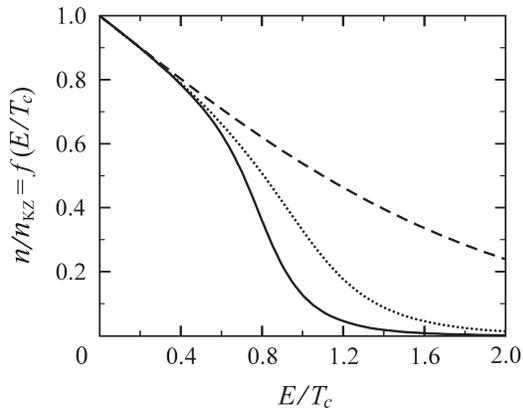}
\caption{\label{fig:DefCon}
Concentration of the defects $n$ normalized to the standard
Kibble--Zurek value $ n_{\rm \scriptscriptstyle KZ} $
as function of the ratio of the domain wall energy $E$ to
the phase transition temperature $T_{c}$ for
the infinite-size 1D Ising model (dashed curve),
$6{\times}6$-cell 2D Ising model with periodic boundary
conditions (solid curve), and
the same model with free boundaries (dotted curve).}
\end{figure}

The general conclusions formulated above can be illustrated
by the particular example in Fig.~\ref{fig:DefCon}, which represents
the concentration of domain walls $n$ normalized
to the standard Kibble--Zurek value $n_{\rm \scriptscriptstyle KZ}$
as function of $ E/T_{c} $ for the following three systems:
\\
(1) one-dimensional Ising model, which admits the exact solution
(see, for example, Ref.~\onlinecite{Isihara71} for more details);
\\
(2) two-dimensional $6{\times}6$-cell Ising model with
\textit{periodic} boundary conditions, which simulates quite well
the system of infinite size; and
\\
(3) two-dimensional $6{\times}6$-cell Ising model with
\textit{free} boundaries (where no energy is concentrated),
which is an example of a microscopic system.
\\
(The particular size of 6 cells along each side of the lattice
was taken quite arbitrarily, just as the value at which
a numerical computation of the statistical sum~(\ref{stat-sum_2D})
is not too cumbersome; details of the calculations can be found
in Appendix.)

%%%%%%%%%% FIGURE 2 %%%%%%%%%%
%%%%%%%%%%%%%%%%%%%%%%%%%%%%%%

As is seen in Fig.~\ref{fig:DefCon}, at $ E/T_{c} \approx 1 $
the concentration of defects in one-dimensional system (dashed curve)
differs from the standard Kibble--Zurek value by less than two times;
in microscopic two-dimensional system (dotted curve), by three times;
while in the macroscopic two-dimensional system (solid curve)
it is suppressed by an order of magnitude. Such suppression becomes
much stronger when the ratio $ E/T_{c} $ increases:
for example, at $ E/T_{c} \approx 2 $ the difference between each of
the curves is over an order of magnitude.

\section{Discussion}

As follows from the above consideration, the specific thermal correlations
between the phases of Bose condensates in separated spatial regions,
revealed for the first time in MJJL experiment, can show us
a general way for explanation of the entire diversity of experimental data
on the topological defect formation. Unfortunately, the results of our
simplified model cannot be confronted quantitatively with the ones
cited in Table~\ref{tab:SumExp}, because the superfluids and
superconductors possess more complex order parameters and
Lagrangians than~(\ref{Lagrangian}). So, more elaborate calculations
are required.

Next, it is important to emphasize that the refined concentration of
defects
\begin{equation}
n = f(E/T_c) \:
  n_{\rm \scriptscriptstyle KZ}({\xi}_\text{eff}%
  ({\tau}_{\scriptscriptstyle Q}))
\end{equation}
possesses exactly the same dependence on
the quench rate ${\tau}_{\scriptscriptstyle Q}$ as in the classical
Kibble--Zurek scenario. So, this dependence, often measured in
the experiments, in general, cannot serve for discrimination between
the models; the absolute values of the defect concentration are
always necessary.

Finally, let us mention that the ideas described in the present work
can be applied also to solving the problem of excessive concentration of
topological defects predicted after the cosmological phase transitions of
Higgs fields.\cite{Dumin03}

\begin{acknowledgments}

One of the authors (Yu.V.\,D.) is grateful to
R.A.~Bertlmann,
Yu.M.~Bunkov,
V.B.~Efimov,
V.B.~Eltsov,
H.J.~Junes,
T.W.B.~Kibble,
M.~Knyazev,
V.P.~Koshelets,
O.D.~Lavrentovich,
A.~Maniv,
P.V.E.~McClintock,
G.R.~Pickett,
E.~Polturak,
A.I.~Rez,
R.J.~Rivers,
A.A.~Starobinsky,
A.V.~Toporensky,
W.G.~Unruh,
D.I.~Uzunov,
G.~Vitiello,
G.E.~Volovik, and
W.H.~Zurek
for valuable discussions and comments.

This work was partially supported by
the European Science Foundation COSLAB Program,
the Abdus Salam International Centre for Theoretical Physics, and
the Institut f{\"u}r Theoretische Physik Karl-Franzens-Universit{\"a}t Graz.

\end{acknowledgments}

\appendix*

\section{Computation of the statistical sums}

For the sake of completeness, let us present the formulas used for drawing
the curves in Fig.~\ref{fig:DefCon}.

%%%%%%%%%% TABLE 3 %%%%%%%%%%
%
\begin{table}
\caption{\label{tab:Coef_Per}
Nonzero statistical weight factors for $6{\times}6$-cell Ising model
with periodic boundary conditions.}
\begin{ruledtabular}
\begin{tabular}{rrrr}
$k$ & $ C^{\text{per}}_{k} $ and $ C^{\text{per}}_{72-k} $ &
$k$ & $ C^{\text{per}}_{k} $ and $ C^{\text{per}}_{72-k} $ \\
\hline
 0 &           2 &   20 &      17\:569\:080 \\
 4 &          72 &   22 &      71\:789\:328 \\
 6 &         144 &   24 &     260\:434\:986 \\
 8 &      1\:620 &   26 &     808\:871\:328 \\
10 &      6\:048 &   28 &  2\:122\:173\:684 \\
12 &     35\:148 &   30 &  4\:616\:013\:408 \\
14 &    159\:840 &   32 &  8\:196\:905\:106 \\
16 &    804\:078 &   34 & 11\:674\:988\:208 \\
18 & 3\:846\:576 &   36 & 13\:172\:279\:424 \\
\end{tabular}
\end{ruledtabular}
\end{table}
%
%%%%%%%%%%%%%%%%%%%%%%%%%%%%%

%%%%%%%%%% TABLE 4 %%%%%%%%%%
%
\begin{table}
\caption{\label{tab:Coef_Free}
Nonzero statistical weight factors for $6{\times}6$-cell Ising model
with free boundaries.}
\begin{ruledtabular}
\begin{tabular}{rrrr}
$k$ & $ C^{\text{free}}_{k} $ and $ C^{\text{free}}_{60-k} $ &
$k$ & $ C^{\text{free}}_{k} $ and $ C^{\text{free}}_{60-k} $ \\
\hline
 0 &           2 &   16 &     15\:444\:302 \\
 2 &           8 &   17 &     33\:435\:520 \\
 3 &          48 &   18 &     69\:487\:240 \\
 4 &         100 &   19 &    138\:380\:976 \\
 5 &         288 &   20 &    263\:185\:168 \\
 6 &      1\:132 &   21 &    476\:852\:512 \\
 7 &      3\:168 &   22 &    821\:190\:292 \\
 8 &      8\:824 &   23 & 1\:340\:056\:928 \\
 9 &     25\:744 &   24 & 2\:065\:952\:532 \\
10 &     71\:064 &   25 & 3\:000\:507\:536 \\
11 &    186\:624 &   26 & 4\:093\:604\:824 \\
12 &    484\:210 &   27 & 5\:230\:849\:920 \\
13 & 1\:214\:336 &   28 & 6\:244\:335\:166 \\
14 & 2\:931\:560 &   29 & 6\:951\:501\:824 \\
15 & 6\:853\:760 &   30 & 7\:206\:345\:520 \\
\end{tabular}
\end{ruledtabular}
\end{table}
%
%%%%%%%%%%%%%%%%%%%%%%%%%%%%%

As is known, all characteristics of the \textit{one-dimensional} Ising model
can be calculated analytically (for a detailed description of the method
see, for example, Ref.~\onlinecite{Isihara71}).
The resulting formula for a defect concentration in a chain of length $N$
with periodic boundary conditions is
\begin{eqnarray}
&& n / n_{\rm \scriptscriptstyle KZ} = \: 2 \, \exp(-E/T_c) \times
\vphantom{\frac{X}{X}}
\\
&& \frac{
   (1 + \exp(-E/T_c))^{N-1} - (1 - \exp(-E/T_c))^{N-1}}{
   (1 + \exp(-E/T_c))^N     + (1 - \exp(-E/T_c))^N} \; ,
\nonumber
\end{eqnarray}
which in the case of infinite chain is reduced to
\begin{equation}
n / n_{\rm \scriptscriptstyle KZ} =
  \frac{2 \, \exp(-E/T_c)}{1 + \exp(-E/T_c)} \; .
\end{equation}

For the \textit{two-dimensional} infinite lattice,
statistical sum~(\ref{gen_stat-sum}) can be reduced by
the well-known technique to some integral over two variables,
and the asymptotics of this integral can be obtained analytically
just in the point of the phase transition of Ising model
(for more details see, for example, Ref.~\onlinecite{Landau69}).
Unfortunately, these results are of little value for our work,
since we need to know the statistical sum in a wide range of temperature.
Besides, this method cannot be applied to the finite-size lattices,
which are also of considerable interest for us.

So, we used a more straightforward approach.
In general, the statistical sum~(\ref{gen_stat-sum}) can be rewritten as
\begin{equation}
Z^{(\text{D})} = \sum_k C_k \exp (-kE /\, T_{c}) \; ,
\label{modif_stat-sum}
\end{equation}
where $ C_k $ is the statistical weight of configurations involving
$k$ elementary domain walls with energy $E$ each.
In the particular case of two-dimensional $6{\times}6$-cell Ising model,
the coefficients $ C_k $ were calculated by the methods of computer algebra
both for the lattice with periodic boundary conditions
(in which every domain wall carries the energy $E$) and
with the free boundaries (where only the inner domain walls contribute
to the total energy of the configuration).
The corresponding coefficients $ C^{\text{per}}_{k} $ and
$ C^{\text{free}}_{k} $ are listed in Tables~\ref{tab:Coef_Per}
and~\ref{tab:Coef_Free}.

Next, the concentration of domain walls is obtained by
substituting~(\ref{modif_stat-sum}) into~(\ref{gen_def-con});
and the final result takes the form
\begin{equation}
n / n_{\rm \scriptscriptstyle KZ} = \: \frac{
  2 \; \displaystyle {\sum}_k k \, C_k \exp(-kE/T_c)}{
  \displaystyle \text{D} N^{\text{D}} \: {\sum}_k C_k \exp(-kE/T_c)} \; ,
\end{equation}
where in the case under consideration $ \text{D} \! = 2 $ and $ N \! = 6 $.


\begin{thebibliography}{18}
\expandafter\ifx\csname natexlab\endcsname\relax\def\natexlab#1{#1}\fi
\expandafter\ifx\csname bibnamefont\endcsname\relax
  \def\bibnamefont#1{#1}\fi
\expandafter\ifx\csname bibfnamefont\endcsname\relax
  \def\bibfnamefont#1{#1}\fi
\expandafter\ifx\csname citenamefont\endcsname\relax
  \def\citenamefont#1{#1}\fi
\expandafter\ifx\csname url\endcsname\relax
  \def\url#1{\texttt{#1}}\fi
\expandafter\ifx\csname urlprefix\endcsname\relax\def\urlprefix{URL }\fi
\providecommand{\bibinfo}[2]{#2}
\providecommand{\eprint}[2][]{\url{#2}}

\bibitem[{\citenamefont{Kibble}(1976)}]{Kibble76}
\bibinfo{author}{\bibfnamefont{T.}~\bibnamefont{Kibble}}, \bibinfo{journal}{J.\
  Phys.\ A} \textbf{\bibinfo{volume}{9}}, \bibinfo{pages}{1387}
  (\bibinfo{year}{1976}).

\bibitem[{\citenamefont{Zurek}(1985)}]{Zurek85}
\bibinfo{author}{\bibfnamefont{W.}~\bibnamefont{Zurek}},
  \bibinfo{journal}{Nature (London)} \textbf{\bibinfo{volume}{317}},
  \bibinfo{pages}{505} (\bibinfo{year}{1985}).

\bibitem[{\citenamefont{Klapdor-Kleingrothaus and Zuber}(1997)}]{Klapdor97}
\bibinfo{author}{\bibfnamefont{H.}~\bibnamefont{Klapdor-Kleingrothaus}}
  \bibnamefont{and} \bibinfo{author}{\bibfnamefont{K.}~\bibnamefont{Zuber}},
  \emph{\bibinfo{title}{Particle Astrophysics}} (\bibinfo{publisher}{Inst.
  Phys. Publ., Bristol}, \bibinfo{year}{1997}).

\bibitem[{\citenamefont{Bogoliubov}(1966)}]{Bogoliubov66}
\bibinfo{author}{\bibfnamefont{N.}~\bibnamefont{Bogoliubov}},
  \bibinfo{journal}{Suppl.\ Nuovo Cimento (Ser.\ prima)}
  \textbf{\bibinfo{volume}{4}}, \bibinfo{pages}{346} (\bibinfo{year}{1966}).

\bibitem[{\citenamefont{Chuang et~al.}(1991)\citenamefont{Chuang, Durrer,
  Turok, and Yurke}}]{Chuang91}
\bibinfo{author}{\bibfnamefont{I.}~\bibnamefont{Chuang}},
  \bibinfo{author}{\bibfnamefont{R.}~\bibnamefont{Durrer}},
  \bibinfo{author}{\bibfnamefont{N.}~\bibnamefont{Turok}}, \bibnamefont{and}
  \bibinfo{author}{\bibfnamefont{B.}~\bibnamefont{Yurke}},
  \bibinfo{journal}{Science} \textbf{\bibinfo{volume}{251}},
  \bibinfo{pages}{1336} (\bibinfo{year}{1991}).

\bibitem[{\citenamefont{Hendry et~al.}(1994)\citenamefont{Hendry, Lawson, Lee,
  McClintock, and Williams}}]{Hendry94}
\bibinfo{author}{\bibfnamefont{P.}~\bibnamefont{Hendry}},
  \bibinfo{author}{\bibfnamefont{N.}~\bibnamefont{Lawson}},
  \bibinfo{author}{\bibfnamefont{R.}~\bibnamefont{Lee}},
  \bibinfo{author}{\bibfnamefont{P.}~\bibnamefont{McClintock}},
  \bibnamefont{and} \bibinfo{author}{\bibfnamefont{C.}~\bibnamefont{Williams}},
  \bibinfo{journal}{Nature (London)} \textbf{\bibinfo{volume}{368}},
  \bibinfo{pages}{315} (\bibinfo{year}{1994}).

\bibitem[{\citenamefont{Dodd et~al.}(1998)\citenamefont{Dodd, Hendry, Lawson,
  McClintock, and Williams}}]{Dodd98}
\bibinfo{author}{\bibfnamefont{M.}~\bibnamefont{Dodd}},
  \bibinfo{author}{\bibfnamefont{P.}~\bibnamefont{Hendry}},
  \bibinfo{author}{\bibfnamefont{N.}~\bibnamefont{Lawson}},
  \bibinfo{author}{\bibfnamefont{P.}~\bibnamefont{McClintock}},
  \bibnamefont{and} \bibinfo{author}{\bibfnamefont{C.}~\bibnamefont{Williams}},
  \bibinfo{journal}{Phys.\ Rev.\ Lett.} \textbf{\bibinfo{volume}{81}},
  \bibinfo{pages}{3703} (\bibinfo{year}{1998}).

\bibitem[{\citenamefont{B{\"a}uerle et~al.}(1996)\citenamefont{B{\"a}uerle,
  Bunkov, Fisher, Godfrin, and Pickett}}]{Bauerle96}
\bibinfo{author}{\bibfnamefont{C.}~\bibnamefont{B{\"a}uerle}},
  \bibinfo{author}{\bibfnamefont{Y.}~\bibnamefont{Bunkov}},
  \bibinfo{author}{\bibfnamefont{S.}~\bibnamefont{Fisher}},
  \bibinfo{author}{\bibfnamefont{H.}~\bibnamefont{Godfrin}}, \bibnamefont{and}
  \bibinfo{author}{\bibfnamefont{G.}~\bibnamefont{Pickett}},
  \bibinfo{journal}{Nature (London)} \textbf{\bibinfo{volume}{382}},
  \bibinfo{pages}{332} (\bibinfo{year}{1996}).

\bibitem[{\citenamefont{Ruutu et~al.}(1996)\citenamefont{Ruutu, Eltsov, Gill,
  Kibble, Krusius, Makhlin, Pla{\c{c}}ais, Volovik, and {Wen Xu}}}]{Ruutu96}
\bibinfo{author}{\bibfnamefont{V.}~\bibnamefont{Ruutu}},
  \bibinfo{author}{\bibfnamefont{V.}~\bibnamefont{Eltsov}},
  \bibinfo{author}{\bibfnamefont{A.}~\bibnamefont{Gill}},
  \bibinfo{author}{\bibfnamefont{T.}~\bibnamefont{Kibble}},
  \bibinfo{author}{\bibfnamefont{M.}~\bibnamefont{Krusius}},
  \bibinfo{author}{\bibfnamefont{Y.}~\bibnamefont{Makhlin}},
  \bibinfo{author}{\bibfnamefont{B.}~\bibnamefont{Pla{\c{c}}ais}},
  \bibinfo{author}{\bibfnamefont{G.}~\bibnamefont{Volovik}}, \bibnamefont{and}
  \bibinfo{author}{\bibnamefont{{Wen Xu}}}, \bibinfo{journal}{Nature (London)}
  \textbf{\bibinfo{volume}{382}}, \bibinfo{pages}{334} (\bibinfo{year}{1996}).

\bibitem[{\citenamefont{Carmi and Polturak}(1999)}]{Carmi99}
\bibinfo{author}{\bibfnamefont{R.}~\bibnamefont{Carmi}} \bibnamefont{and}
  \bibinfo{author}{\bibfnamefont{E.}~\bibnamefont{Polturak}},
  \bibinfo{journal}{Phys.\ Rev.\ B} \textbf{\bibinfo{volume}{60}},
  \bibinfo{pages}{7595} (\bibinfo{year}{1999}).

\bibitem[{\citenamefont{Maniv et~al.}(2003)\citenamefont{Maniv, Polturak, and
  Koren}}]{Maniv03}
\bibinfo{author}{\bibfnamefont{A.}~\bibnamefont{Maniv}},
  \bibinfo{author}{\bibfnamefont{E.}~\bibnamefont{Polturak}}, \bibnamefont{and}
  \bibinfo{author}{\bibfnamefont{G.}~\bibnamefont{Koren}},
  \bibinfo{journal}{Phys.\ Rev.\ Lett.} \textbf{\bibinfo{volume}{91}},
  \bibinfo{pages}{197001} (\bibinfo{year}{2003}).

\bibitem[{\citenamefont{Kirtley et~al.}(2003)\citenamefont{Kirtley, Tsuei, and
  Tafuri}}]{Kirtley03}
\bibinfo{author}{\bibfnamefont{J.}~\bibnamefont{Kirtley}},
  \bibinfo{author}{\bibfnamefont{C.}~\bibnamefont{Tsuei}}, \bibnamefont{and}
  \bibinfo{author}{\bibfnamefont{F.}~\bibnamefont{Tafuri}},
  \bibinfo{journal}{Phys.\ Rev.\ Lett.} \textbf{\bibinfo{volume}{90}},
  \bibinfo{pages}{257001} (\bibinfo{year}{2003}).

\bibitem[{\citenamefont{Carmi et~al.}(2000)\citenamefont{Carmi, Polturak, and
  Koren}}]{Carmi00}
\bibinfo{author}{\bibfnamefont{R.}~\bibnamefont{Carmi}},
  \bibinfo{author}{\bibfnamefont{E.}~\bibnamefont{Polturak}}, \bibnamefont{and}
  \bibinfo{author}{\bibfnamefont{G.}~\bibnamefont{Koren}},
  \bibinfo{journal}{Phys.\ Rev.\ Lett.} \textbf{\bibinfo{volume}{84}},
  \bibinfo{pages}{4966} (\bibinfo{year}{2000}).

\bibitem[{\citenamefont{Monaco et~al.}(2002)\citenamefont{Monaco, Mygind, and
  Rivers}}]{Monaco02}
\bibinfo{author}{\bibfnamefont{R.}~\bibnamefont{Monaco}},
  \bibinfo{author}{\bibfnamefont{J.}~\bibnamefont{Mygind}}, \bibnamefont{and}
  \bibinfo{author}{\bibfnamefont{R.}~\bibnamefont{Rivers}},
  \bibinfo{journal}{Phys.\ Rev.\ Lett.} \textbf{\bibinfo{volume}{89}},
  \bibinfo{pages}{080603} (\bibinfo{year}{2002}).

\bibitem[{\citenamefont{Rumer and Ryvkin}(1980)}]{Rumer80}
\bibinfo{author}{\bibfnamefont{Y.}~\bibnamefont{Rumer}} \bibnamefont{and}
  \bibinfo{author}{\bibfnamefont{M.}~\bibnamefont{Ryvkin}},
  \emph{\bibinfo{title}{Thermodynamics, Statistical Physics, and Kinetics}}
  (\bibinfo{publisher}{Mir, Moscow}, \bibinfo{year}{1980}).

\bibitem[{\citenamefont{Isihara}(1971)}]{Isihara71}
\bibinfo{author}{\bibfnamefont{A.}~\bibnamefont{Isihara}},
  \emph{\bibinfo{title}{Statistical Physics}} (\bibinfo{publisher}{Academic
  Press, NY}, \bibinfo{year}{1971}).

\bibitem[{\citenamefont{Dumin}(2003)}]{Dumin03}
\bibinfo{author}{\bibfnamefont{Y.}~\bibnamefont{Dumin}}, in
  \emph{\bibinfo{booktitle}{Frontiers of Particle Physics}}
  (\bibinfo{publisher}{World Scientific Co, Singapore}, \bibinfo{year}{2003}),
  p. \bibinfo{pages}{289}, \eprint{hep-ph/0204154}.

\bibitem[{\citenamefont{Landau and Lifshitz}(1969)}]{Landau69}
\bibinfo{author}{\bibfnamefont{L.}~\bibnamefont{Landau}} \bibnamefont{and}
  \bibinfo{author}{\bibfnamefont{E.}~\bibnamefont{Lifshitz}},
  \emph{\bibinfo{title}{Statistical Physics}} (\bibinfo{publisher}{Pergamon
  Press, Oxford}, \bibinfo{year}{1969}).

\end{thebibliography}
\end{document}